\documentclass[conference,10pt,twoside,twocolumn]{IEEEtran}

\usepackage{mleftright}
\usepackage{cite}
\usepackage[T1]{fontenc}
\usepackage{graphicx}
\usepackage{mathtools}
\usepackage{amssymb}
\usepackage{amsmath}
\usepackage{amsthm}

\usepackage{paralist}
\usepackage{subfigure}
\usepackage{booktabs} %\toprule \midrule \bottomrule
\usepackage{algorithm}
\usepackage{multirow}
\usepackage{microtype}
\usepackage{tablefootnote}
\usepackage{balance}
\usepackage{hyperref}

\usepackage{amssymb}
\usepackage{amsfonts}
\usepackage{mathrsfs}
\usepackage{xspace}
\usepackage{bm}
\usepackage{upgreek}

\newcommand{\safemath}[2]{\newcommand{#1}{\ensuremath{#2}\xspace}}

\safemath{\bma}{\mathbf{a}}
\safemath{\bmb}{\mathbf{b}}
\safemath{\bmc}{\mathbf{c}}
\safemath{\bmd}{\mathbf{d}}
\safemath{\bme}{\mathbf{e}}
\safemath{\bmf}{\mathbf{f}}
\safemath{\bmg}{\mathbf{g}}
\safemath{\bmh}{\mathbf{h}}
\safemath{\bmi}{\mathbf{i}}
\safemath{\bmj}{\mathbf{j}}
\safemath{\bmk}{\mathbf{k}}
\safemath{\bml}{\mathbf{l}}
\safemath{\bmm}{\mathbf{m}}
\safemath{\bmn}{\mathbf{n}}
\safemath{\bmo}{\mathbf{o}}
\safemath{\bmp}{\mathbf{p}}
\safemath{\bmq}{\mathbf{q}}
\safemath{\bmr}{\mathbf{r}}
\safemath{\bms}{\mathbf{s}}
\safemath{\bmt}{\mathbf{t}}
\safemath{\bmu}{\mathbf{u}}
\safemath{\bmv}{\mathbf{v}}
\safemath{\bmw}{\mathbf{w}}
\safemath{\bmx}{\mathbf{x}}
\safemath{\bmy}{\mathbf{y}}
\safemath{\bmz}{\mathbf{z}}
\safemath{\bmzero}{\mathbf{0}}
\safemath{\bmone}{\mathbf{1}}

\bmdefine{\biad}{a}
\bmdefine{\bibd}{b}
\bmdefine{\bicd}{c}
\bmdefine{\bidd}{d}
\bmdefine{\bied}{e}
\bmdefine{\bifd}{f}
\bmdefine{\bigd}{g}
\bmdefine{\bihd}{h}
\bmdefine{\biid}{i}
\bmdefine{\bijd}{j}
\bmdefine{\bikd}{k}
\bmdefine{\bild}{l}
\bmdefine{\bimd}{m}
\bmdefine{\bind}{n}
\bmdefine{\biod}{o}
\bmdefine{\bipd}{p}
\bmdefine{\biqd}{q}
\bmdefine{\bird}{r}
\bmdefine{\bisd}{s}
\bmdefine{\bitd}{t}
\bmdefine{\biud}{u}
\bmdefine{\bivd}{v}
\bmdefine{\biwd}{w}
\bmdefine{\bixd}{x}
\bmdefine{\biyd}{y}
\bmdefine{\bizd}{z}

\bmdefine{\bixid}{\xi}
\bmdefine{\bilambdad}{\lambda}
\bmdefine{\bimud}{\mu}
\bmdefine{\bithetad}{\theta}
\bmdefine{\biphid}{\phi}
\bmdefine{\bideltad}{\delta}

\safemath{\bmia}{\biad}
\safemath{\bmib}{\bibd}
\safemath{\bmic}{\bicd}
\safemath{\bmid}{\bidd}
\safemath{\bmie}{\bied}
\safemath{\bmif}{\bifd}
\safemath{\bmig}{\bigd}
\safemath{\bmih}{\bihd}
\safemath{\bmii}{\biid}
\safemath{\bmij}{\bijd}
\safemath{\bmik}{\bikd}
\safemath{\bmil}{\bild}
\safemath{\bmim}{\bimd}
\safemath{\bmin}{\bind}
\safemath{\bmio}{\biod}
\safemath{\bmip}{\bipd}
\safemath{\bmiq}{\biqd}
\safemath{\bmir}{\bird}
\safemath{\bmis}{\bisd}
\safemath{\bmit}{\bitd}
\safemath{\bmiu}{\biud}
\safemath{\bmiv}{\bivd}
\safemath{\bmiw}{\biwd}
\safemath{\bmix}{\bixd}
\safemath{\bmiy}{\biyd}
\safemath{\bmiz}{\bizd}

\safemath{\bmxi}{\bixid}
\safemath{\bmlambda}{\bilambdad}
\safemath{\bmmu}{\bimud}
\safemath{\bmtheta}{\bithetad}
\safemath{\bmphi}{\biphid}
\safemath{\bmdelta}{\bideltad}

\safemath{\bA}{\mathbf{A}}
\safemath{\bB}{\mathbf{B}}
\safemath{\bC}{\mathbf{C}}
\safemath{\bD}{\mathbf{D}}
\safemath{\bE}{\mathbf{E}}
\safemath{\bF}{\mathbf{F}}
\safemath{\bG}{\mathbf{G}}
\safemath{\bH}{\mathbf{H}}
\safemath{\bI}{\mathbf{I}}
\safemath{\bJ}{\mathbf{J}}
\safemath{\bK}{\mathbf{K}}
\safemath{\bL}{\mathbf{L}}
\safemath{\bM}{\mathbf{M}}
\safemath{\bN}{\mathbf{N}}
\safemath{\bO}{\mathbf{O}}
\safemath{\bP}{\mathbf{P}}
\safemath{\bQ}{\mathbf{Q}}
\safemath{\bR}{\mathbf{R}}
\safemath{\bS}{\mathbf{S}}
\safemath{\bT}{\mathbf{T}}
\safemath{\bU}{\mathbf{U}}
\safemath{\bV}{\mathbf{V}}
\safemath{\bW}{\mathbf{W}}
\safemath{\bX}{\mathbf{X}}
\safemath{\bY}{\mathbf{Y}}
\safemath{\bZ}{\mathbf{Z}}

\safemath{\bZero}{\mathbf{0}}
\safemath{\bOne}{\mathbf{1}}
\safemath{\bDelta}{\mathbf{\Delta}}
\safemath{\bLambda}{\mathbf{\UpLambda}}
\safemath{\bPhi}{\mathbf{\Upphi}}
\safemath{\bSigma}{\mathbf{\Upsigma}}
\safemath{\bOmega}{\mathbf{\Upomega}}
\safemath{\bTheta}{\mathbf{\Uptheta}}

\bmdefine{\biAd}{A}
\bmdefine{\biBd}{B}
\bmdefine{\biCd}{C}
\bmdefine{\biDd}{D}
\bmdefine{\biEd}{E}
\bmdefine{\biFd}{F}
\bmdefine{\biGd}{G}
\bmdefine{\biHd}{H}
\bmdefine{\biId}{I}
\bmdefine{\biJd}{J}
\bmdefine{\biKd}{K}
\bmdefine{\biLd}{L}
\bmdefine{\biMd}{M}
\bmdefine{\biOd}{N}
\bmdefine{\biPd}{O}
\bmdefine{\biQd}{P}
\bmdefine{\biRd}{R}
\bmdefine{\biSd}{S}
\bmdefine{\biTd}{T}
\bmdefine{\biUd}{U}
\bmdefine{\biVd}{V}
\bmdefine{\biWd}{W}
\bmdefine{\biXd}{X}
\bmdefine{\biYd}{Y}
\bmdefine{\biZd}{Z}

\bmdefine{\biDelta}{\Delta}
\bmdefine{\biLambda}{\Lambda}
\bmdefine{\biPhi}{\Phi}
\bmdefine{\biSigma}{\Sigma}
\bmdefine{\biOmega}{\Omega}
\bmdefine{\biTheta}{\Theta}

\safemath{\bimA}{\biAd}
\safemath{\bimB}{\biBd}
\safemath{\bimC}{\biCd}
\safemath{\bimD}{\biDd}
\safemath{\bimE}{\biEd}
\safemath{\bimF}{\biFd}
\safemath{\bimG}{\biGd}
\safemath{\bimH}{\biHd}
\safemath{\bimI}{\biId}
\safemath{\bimJ}{\biJd}
\safemath{\bimK}{\biKd}
\safemath{\bimL}{\biLd}
\safemath{\bimM}{\biMd}
\safemath{\bimN}{\biNd}
\safemath{\bimO}{\biOd}
\safemath{\bimP}{\biPd}
\safemath{\bimQ}{\biQd}
\safemath{\bimR}{\biRd}
\safemath{\bimS}{\biSd}
\safemath{\bimT}{\biTd}
\safemath{\bimU}{\biUd}
\safemath{\bimV}{\biVd}
\safemath{\bimW}{\biWd}
\safemath{\bimX}{\biXd}
\safemath{\bimY}{\biYd}
\safemath{\bimZ}{\biZd}

\safemath{\bimDelta}{\biDelta}
\safemath{\bimLambda}{\biLambda}
\safemath{\bimPhi}{\biPhi}
\safemath{\bimSigma}{\biSigma}
\safemath{\bimOmega}{\biOmega}
\safemath{\bimTheta}{\biTheta}

\safemath{\setA}{\mathcal{A}}
\safemath{\setB}{\mathcal{B}}
\safemath{\setC}{\mathcal{C}}
\safemath{\setD}{\mathcal{D}}
\safemath{\setE}{\mathcal{E}}
\safemath{\setF}{\mathcal{F}}
\safemath{\setG}{\mathcal{G}}
\safemath{\setH}{\mathcal{H}}
\safemath{\setI}{\mathcal{I}}
\safemath{\setJ}{\mathcal{J}}
\safemath{\setK}{\mathcal{K}}
\safemath{\setL}{\mathcal{L}}
\safemath{\setM}{\mathcal{M}}
\safemath{\setN}{\mathcal{N}}
\safemath{\setO}{\mathcal{O}}
\safemath{\setP}{\mathcal{P}}
\safemath{\setQ}{\mathcal{Q}}
\safemath{\setR}{\mathcal{R}}
\safemath{\setS}{\mathcal{S}}
\safemath{\setT}{\mathcal{T}}
\safemath{\setU}{\mathcal{U}}
\safemath{\setV}{\mathcal{V}}
\safemath{\setW}{\mathcal{W}}
\safemath{\setX}{\mathcal{X}}
\safemath{\setY}{\mathcal{Y}}
\safemath{\setZ}{\mathcal{Z}}
\safemath{\emptySet}{\varnothing}

\safemath{\colA}{\mathscr{A}}
\safemath{\colB}{\mathscr{B}}
\safemath{\colC}{\mathscr{C}}
\safemath{\colD}{\mathscr{D}}
\safemath{\colE}{\mathscr{E}}
\safemath{\colF}{\mathscr{F}}
\safemath{\colG}{\mathscr{G}}
\safemath{\colH}{\mathscr{H}}
\safemath{\colI}{\mathscr{I}}
\safemath{\colJ}{\mathscr{J}}
\safemath{\colK}{\mathscr{K}}
\safemath{\colL}{\mathscr{L}}
\safemath{\colM}{\mathscr{M}}
\safemath{\colN}{\mathscr{N}}
\safemath{\colO}{\mathscr{O}}
\safemath{\colP}{\mathscr{P}}
\safemath{\colQ}{\mathscr{Q}}
\safemath{\colR}{\mathscr{R}}
\safemath{\colS}{\mathscr{S}}
\safemath{\colT}{\mathscr{T}}
\safemath{\colU}{\mathscr{U}}
\safemath{\colV}{\mathscr{V}}
\safemath{\colW}{\mathscr{W}}
\safemath{\colX}{\mathscr{X}}
\safemath{\colY}{\mathscr{Y}}
\safemath{\colZ}{\mathscr{Z}}

\safemath{\opA}{\mathbb{A}}
\safemath{\opB}{\mathbb{B}}
\safemath{\opC}{\mathbb{C}}
\safemath{\opD}{\mathbb{D}}
\safemath{\opE}{\mathbb{E}}
\safemath{\opF}{\mathbb{F}}
\safemath{\opG}{\mathbb{G}}
\safemath{\opH}{\mathbb{H}}
\safemath{\opI}{\mathbb{I}}
\safemath{\opJ}{\mathbb{J}}
\safemath{\opK}{\mathbb{K}}
\safemath{\opL}{\mathbb{L}}
\safemath{\opM}{\mathbb{M}}
\safemath{\opN}{\mathbb{N}}
\safemath{\opO}{\mathbb{O}}
\safemath{\opP}{\mathbb{P}}
\safemath{\opQ}{\mathbb{Q}}
\safemath{\opR}{\mathbb{R}}
\safemath{\opS}{\mathbb{S}}
\safemath{\opT}{\mathbb{T}}
\safemath{\opU}{\mathbb{U}}
\safemath{\opV}{\mathbb{V}}
\safemath{\opW}{\mathbb{W}}
\safemath{\opX}{\mathbb{X}}
\safemath{\opY}{\mathbb{Y}}
\safemath{\opZ}{\mathbb{Z}}
\safemath{\opZero}{\mathbb{O}}
\safemath{\identityop}{\opI}

\safemath{\veca}{\bma}
\safemath{\vecb}{\bmb}
\safemath{\vecc}{\bmc}
\safemath{\vecd}{\bmd}
\safemath{\vece}{\bme}
\safemath{\vecf}{\bmf}
\safemath{\vecg}{\bmg}
\safemath{\vech}{\bmh}
\safemath{\veci}{\bmi}
\safemath{\vecj}{\bmj}
\safemath{\veck}{\bmk}
\safemath{\vecl}{\bml}
\safemath{\vecm}{\bmm}
\safemath{\vecn}{\bmn}
\safemath{\veco}{\bmo}
\safemath{\vecp}{\bmp}
\safemath{\vecq}{\bmq}
\safemath{\vecr}{\bmr}
\safemath{\vecs}{\bms}
\safemath{\vect}{\bmt}
\safemath{\vecu}{\bmu}
\safemath{\vecv}{\bmv}
\safemath{\vecw}{\bmw}
\safemath{\vecx}{\bmx}
\safemath{\vecy}{\bmy}
\safemath{\vecz}{\bmz}

\safemath{\veczero}{\bmzero}
\safemath{\vecone}{\bmone}
\safemath{\vecxi}{\bmxi}
\safemath{\veclambda}{\bmlambda}
\safemath{\vecmu}{\bmmu}
\safemath{\vectheta}{\bmtheta}
\safemath{\vecphi}{\bmphi}
\safemath{\vecdelta}{\bmdelta}

\safemath{\matA}{\bA}
\safemath{\matB}{\bB}
\safemath{\matC}{\bC}
\safemath{\matD}{\bD}
\safemath{\matE}{\bE}
\safemath{\matF}{\bF}
\safemath{\matG}{\bG}
\safemath{\matH}{\bH}
\safemath{\matI}{\bI}
\safemath{\matJ}{\bJ}
\safemath{\matK}{\bK}
\safemath{\matL}{\bL}
\safemath{\matM}{\bM}
\safemath{\matN}{\bN}
\safemath{\matO}{\bO}
\safemath{\matP}{\bP}
\safemath{\matQ}{\bQ}
\safemath{\matR}{\bR}
\safemath{\matS}{\bS}
\safemath{\matT}{\bT}
\safemath{\matU}{\bU}
\safemath{\matV}{\bV}
\safemath{\matW}{\bW}
\safemath{\matX}{\bX}
\safemath{\matY}{\bY}
\safemath{\matZ}{\bZ}
\safemath{\matzero}{\bmzero}

\safemath{\matDelta}{\bDelta}
\safemath{\matLambda}{\bLambda}
\safemath{\matPhi}{\bPhi}
\safemath{\matSigma}{\bSigma}
\safemath{\matOmega}{\bOmega}
\safemath{\matTheta}{\bTheta}

\safemath{\matidentity}{\matI}
\safemath{\matone}{\matO}

\safemath{\rnda}{A}
\safemath{\rndb}{B}
\safemath{\rndc}{C}
\safemath{\rndd}{D}
\safemath{\rnde}{E}
\safemath{\rndf}{F}
\safemath{\rndg}{G}
\safemath{\rndh}{H}
\safemath{\rndi}{I}
\safemath{\rndj}{J}
\safemath{\rndk}{K}
\safemath{\rndl}{L}
\safemath{\rndm}{M}
\safemath{\rndn}{N}
\safemath{\rndo}{O}
\safemath{\rndp}{P}
\safemath{\rndq}{Q}
\safemath{\rndr}{R}
\safemath{\rnds}{S}
\safemath{\rndt}{T}
\safemath{\rndu}{U}
\safemath{\rndv}{V}
\safemath{\rndw}{W}
\safemath{\rndx}{X}
\safemath{\rndy}{Y}
\safemath{\rndz}{Z}

\safemath{\rveca}{\bimA}
\safemath{\rvecb}{\bimB}
\safemath{\rvecc}{\bimC}
\safemath{\rvecd}{\bimD}
\safemath{\rvece}{\bimE}
\safemath{\rvecf}{\bimF}
\safemath{\rvecg}{\bimG}
\safemath{\rvech}{\bimH}
\safemath{\rveci}{\bimI}
\safemath{\rvecj}{\bimJ}
\safemath{\rveck}{\bimK}
\safemath{\rvecl}{\bimL}
\safemath{\rvecm}{\bimM}
\safemath{\rvecn}{\bimN}
\safemath{\rveco}{\bomO}
\safemath{\rvecp}{\bimP}
\safemath{\rvecq}{\bimQ}
\safemath{\rvecr}{\bimR}
\safemath{\rvecs}{\bimS}
\safemath{\rvect}{\bimT}
\safemath{\rvecu}{\bimU}
\safemath{\rvecv}{\bimV}
\safemath{\rvecw}{\bimW}
\safemath{\rvecx}{\bimX}
\safemath{\rvecy}{\bimY}
\safemath{\rvecz}{\bimZ}

\safemath{\rvecxi}{\bmxi}
\safemath{\rveclambda}{\bmlambda}
\safemath{\rvecmu}{\bmmu}
\safemath{\rvectheta}{\bmtheta}
\safemath{\rvecphi}{\bmphi}

\safemath{\rmatA}{\bimA}
\safemath{\rmatB}{\bimB}
\safemath{\rmatC}{\bimC}
\safemath{\rmatD}{\bimD}
\safemath{\rmatE}{\bimE}
\safemath{\rmatF}{\bimF}
\safemath{\rmatG}{\bimG}
\safemath{\rmatH}{\bimH}
\safemath{\rmatI}{\bimI}
\safemath{\rmatJ}{\bimJ}
\safemath{\rmatK}{\bimK}
\safemath{\rmatL}{\bimL}
\safemath{\rmatM}{\bimM}
\safemath{\rmatN}{\bimN}
\safemath{\rmatO}{\bimO}
\safemath{\rmatP}{\bimP}
\safemath{\rmatQ}{\bimQ}
\safemath{\rmatR}{\bimR}
\safemath{\rmatS}{\bimS}
\safemath{\rmatT}{\bimT}
\safemath{\rmatU}{\bimU}
\safemath{\rmatV}{\bimV}
\safemath{\rmatW}{\bimW}
\safemath{\rmatX}{\bimX}
\safemath{\rmatY}{\bimY}
\safemath{\rmatZ}{\bimZ}

\safemath{\rmatDelta}{\bimDelta}
\safemath{\rmatLambda}{\bimLambda}
\safemath{\rmatPhi}{\bimPhi}
\safemath{\rmatSigma}{\bimSigma}
\safemath{\rmatOmega}{\bimOmega}
\safemath{\rmatTheta}{\bimTheta}

\usepackage{amssymb}
\usepackage{amsfonts}
\usepackage{mathrsfs}
\usepackage{xspace}
\usepackage{bm}
\usepackage{fancyref}
\usepackage{textcomp}

\usepackage{multirow}
\usepackage{stmaryrd}

\newenvironment{textbmatrix}{	\setlength{\arraycolsep}{2.5pt}%
								\big[\begin{matrix}}{\end{matrix}\big]%
								\raisebox{0.08ex}{\vphantom{M}}}

\def\be{\begin{equation}}
\def\ee{\end{equation}}
\def\een{\nonumber \end{equation}}
\def\mat{\begin{bmatrix}}
\def\emat{\end{bmatrix}}
\def\btm{\begin{textbmatrix}}
\def\etm{\end{textbmatrix}}

\def\ba#1\ea{\begin{align}#1\end{align}}
\def\bas#1\eas{\begin{align*}#1\end{align*}}
\def\bs#1\es{\begin{split}#1\end{split}}
\def\bg#1\eg{\begin{gather}#1\end{gather}}
\def\bml#1\eml{\begin{multline}#1\end{multline}}
\def\bi#1\ei{\begin{itemize}#1\end{itemize}}

\safemath{\dirac}{\delta}					% Dirac delta
\safemath{\krond}{\dirac}					% Kronecker delta
\safemath{\upto}{\uparrow}
\safemath{\downto}{\downarrow}
\safemath{\iu}{j}							% imaginary unit
\safemath{\ev}{\lambda}						% eigenvalue
\safemath{\hilseqspace}{l^{2}}				% Hilbert sequence space
\newcommand{\banachfunspace}[1]{\setL^{#1}}	% Banach function space
\safemath{\hilfunspace}{\banachfunspace{2}}	% Hilbert function space
\newcommand{\ceil}[1]{\lceil #1 \rceil}

\safemath{\SNR}{\textit{SNR}} 				% signal to noise ratio
\safemath{\PAR}{\textit{PAR}} 				% signal to noise ratio
\safemath{\No}{N_0}							% noise spectral density
\safemath{\Es}{E_s}							% energy per symbol
\safemath{\Eb}{E_b}							% energy per bit
\safemath{\EbNo}{\frac{\Eb}{\No}}
\safemath{\EsNo}{\frac{\Es}{\No}}

\DeclareMathOperator{\CHop}{\ensuremath{\opH}} % channel operator
\safemath{\tvir}{\rndh_{\CHop}}				% time-varying impulse response
\safemath{\tvtf}{\rndl_{\CHop}}				% 	-''- transfer function
\safemath{\spf}{\rnds_{\CHop}}				% spreading function
\safemath{\bff}{H_{\CHop}}					% bi-freuqency function

\safemath{\ircf}{r_{h}}						% impulse response correlation fn.
\safemath{\tftvcf}{r_{s}}					% scattering function
\safemath{\tfcf}{r_{l}}						% time-frequency correlation fn.
\safemath{\bfcf}{r_{H}}						% bi-frequency correlation fn.

\safemath{\tcorr}{c_h}						% time-correlation function
\safemath{\scf}{c_{s}}						% spreading function
\safemath{\tfcorr}{c_{l}}					% transfer-function correlation
\safemath{\fcorr}{c_{H}}						% frequency-correlation function

\safemath{\mi}{I}							% mutual information
\safemath{\capacity}{C}						% capacity

\safemath{\normal}{\mathcal{N}}			% normal distribution
\safemath{\jpg}{\mathcal{CN}}			% jointly proper Gaussian
\safemath{\mchain}{\leftrightarrow}		% Markov chain
\safemath{\dB}{\,\mathrm{dB}}
\safemath{\dBm}{\,\mathrm{dBm}}
\safemath{\Hz}{\,\mathrm{Hz}}
\safemath{\kHz}{\,\mathrm{kHz}}
\safemath{\MHz}{\,\mathrm{MHz}}
\safemath{\GHz}{\,\mathrm{GHz}}
\safemath{\s}{\,\mathrm{s}}
\safemath{\ms}{\,\mathrm{ms}}
\safemath{\mus}{\,\mathrm{\text{\textmu}s}}
\safemath{\ns}{\,\mathrm{ns}}
\safemath{\ps}{\,\mathrm{ps}}
\safemath{\meter}{\,\mathrm{m}}
\safemath{\mm}{\,\mathrm{mm}}
\safemath{\cm}{\,\mathrm{cm}}
\safemath{\m}{\,\mathrm{m}}
\safemath{\W}{\,\mathrm{W}}
\safemath{\mW}{\, \mathrm{mW}}
\safemath{\J}{\,\mathrm{J}}
\safemath{\K}{\,\mathrm{K}}
\safemath{\bit}{\,\mathrm{bit}}
\safemath{\nat}{\,\mathrm{nat}}

\safemath{\define}{\triangleq}			% definition

\safemath{\equivalent}{\sim}
\safemath{\distas}{\sim}					% distributed according to
\safemath{\sdiff}{\Delta}				% symmetric set difference

\safemath{\reals}{\mathbb{R}}
\safemath{\positivereals}{\reals_{+}}
\safemath{\integers}{\mathbb{Z}}
\safemath{\posint}{\integers_{+}}
\safemath{\naturals}{\mathbb{N}}
\safemath{\posnaturals}{\naturals_{+}}
\safemath{\complexset}{\mathbb{C}}
\safemath{\rationals}{\mathbb{Q}}

\newcommand*{\fancyrefapplabelprefix}{app}		% Appendix
\newcommand*{\fancyrefthmlabelprefix}{thm}		% Theorem
\newcommand*{\fancyreflemlabelprefix}{lem}		% Lemma
\newcommand*{\fancyrefcorlabelprefix}{cor}		% Corollary
\newcommand*{\fancyrefdeflabelprefix}{def}		% Definition
\newcommand*{\fancyrefproplabelprefix}{prop}		% Proposition
\newcommand*{\fancyrefexmpllabelprefix}{exmpl}
\newcommand*{\fancyrefalglabelprefix}{alg}		% Algorithm
\newcommand*{\fancyreftbllabelprefix}{tbl}		% Algorithm

\frefformat{vario}{\fancyrefseclabelprefix}{Section~#1}
\frefformat{vario}{\fancyrefthmlabelprefix}{Theorem~#1}
\frefformat{vario}{\fancyreftbllabelprefix}{Table~#1}
\frefformat{vario}{\fancyreflemlabelprefix}{Lemma~#1}
\frefformat{vario}{\fancyrefcorlabelprefix}{Corollary~#1}
\frefformat{vario}{\fancyrefdeflabelprefix}{Definition~#1}
\frefformat{vario}{\fancyreffiglabelprefix}{Fig.~#1}
\frefformat{vario}{\fancyrefapplabelprefix}{Appendix~#1}
\frefformat{vario}{\fancyrefeqlabelprefix}{(#1)}
\frefformat{vario}{\fancyrefproplabelprefix}{Proposition~#1}
\frefformat{vario}{\fancyrefexmpllabelprefix}{Example~#1}
\frefformat{vario}{\fancyrefalglabelprefix}{Algorithm~#1}

\safemath{\dictab}{[\,\dicta\,\,\dictb\,]}

\safemath{\ysig}{\bmy}
\safemath{\ysighat}{\hat{\ysig}}
\safemath{\ysigdim}{M}
\safemath{\xsig}{\bmx}
\safemath{\xsigdim}{N}
\safemath{\nx}{n_x}
\safemath{\zsig}{\bmz}
\safemath{\zsigdim}{\ysigdim}
\safemath{\rsig}{\bmr}
\safemath{\Adict}{\bA}
\safemath{\Adicttilde}{\widetilde{\Adict}}
\safemath{\Adictdim}{\outputdim\times\xsigdim}
\safemath{\avec}{\bma}
\safemath{\avectilde}{\tilde{\avec}}
\safemath{\Bdict}{\bB}
\safemath{\Bdicttilde}{\widetilde{\Bdict}}
\safemath{\Cdict}{\bC}
\safemath{\cvec}{\bmc}
\safemath{\Ddict}{\bD}
\safemath{\Ddictdim}{\ysigdim\times\xsigdim}
\safemath{\dvec}{\bmd}
\safemath{\Ddicttilde}{\widetilde{\bD}}
\safemath{\Bonb}{\bB}
\safemath{\bvec}{\bmb}
\safemath{\Bonbdim}{\ysigdim\times\ysigdim}
\safemath{\noise}{\bmn}
\safemath{\noisedim}{\ysigim}
\safemath{\err}{\bme}
\safemath{\errdim}{\ysigdim}
\safemath{\errset}{\setE}
\safemath{\nerr}{n_e}
\safemath{\delop}{\bP_\errset}
\safemath{\delopc}{\bP_{{\errset}^c}}

\safemath{\cplxi}{\imath}
\safemath{\cplxj}{\jmath}
\safemath{\dict}{\matD}
\safemath{\inputdim}{N}		% number of columns of dictionary D
\safemath{\outputdim}{M}		%number of rows of dictionary D
\safemath{\sparsity}{S}	%sparsity
\safemath{\inputdimA}{{N_a}}	%total number of elements in dictionary A
\safemath{\inputdimB}{{N_b}}	%total number of elements in dictionary B
\safemath{\elemA}{{n_a}}	%number of elements chosen from dictionary A
\safemath{\elemB}{{n_b}}	%number of elements chosen from dictionary B
\safemath{\resA}{\matR_a}	%restriction map to elements of dictionary A
\safemath{\resB}{\matR_b}	%restriction map to elements of dictionary B
\safemath{\subD}{\matS} %subdictionary
\safemath{\subA}{\matS_a} %subdictionary part of A
\safemath{\subB}{\matS_b} %subdictionary part of B
\safemath{\dicta}{\matA} 	% first subdictionary
\safemath{\dictb}{\matB} 	% second subdictionary
\safemath{\hollowS}{H}
\safemath{\hollowA}{H_a}
\safemath{\hollowB}{H_b}
\safemath{\cross}{Z}
\safemath{\coh}{\mu_d}			% coherence dictionary
\safemath{\coha}{\mu_a}			% coherence first subdictionary
\safemath{\cohb}{\mu_b}			% coherence second subdictionary
\safemath{\mubs}{\nu}	%block sub-coherence
\safemath{\cohm}{\mu_m} %mutual coherence
\safemath{\dictset}{\setD}	% set of dictionaries
\safemath{\dictsetp}{\dictset(\coh,\coha,\cohb)}	% set of dictionaries parametrized
\safemath{\dictsetgen}{\dictset_\text{gen}}
\safemath{\dictsetgenp}{\dictsetgen(\coh)}
\safemath{\dictsetonb}{\dictset_\text{onb}}
\safemath{\dictsetonbp}{\dictsetonb(\coh)}

\safemath{\leftside}{U}
\safemath{\rightsideA}{R_a}
\safemath{\rightsideB}{R_b}

\safemath{\indexS}{\setI_S} %set of indices participating in sub-dictionary S

\safemath{\na}{n_a}			% cardinality of set of linearly independent columns of first dictionary
\safemath{\nb}{n_b}			% cardinality of set of linearly independent columns of second dictionary
\safemath{\coeffa}{p_i}	%coefficients for columns of A
\safemath{\coeffb}{q_j}	%coefficients for columns of B
\safemath{\seta}{\setP}		% set of linearly independent columns of A
\safemath{\setb}{\setQ}     % set of linearly independent columns of B
\safemath{\setw}{\setW}	%set of n largest elements of w
\safemath{\setz}{\setZ}	%set of L-n largest elements of z
\safemath{\cola}{\veca}		% generic element of the dictionary A
\safemath{\colb}{\vecb}		% generic element of the dictionary B
\safemath{\cold}{\vecd}		% generic element of the dictionary D
\safemath{\inputvec}{\vecx} 	%coefficient vector (input)
\safemath{\error}{\vece}	%error vector
\safemath{\noiseout}{\vecz} 	%noisy output vector
\safemath{\inputvecel}{x}
\safemath{\inputveca}{\vecx_a}
\safemath{\inputvecb}{\vecx_b}
\safemath{\outputvec}{\vecy}	%output of Dictionary
\safemath{\lambdamin}{\lambda_{\mathrm{min}}}
\safemath{\elltwo}{\ell_2}
\safemath{\ellone}{\ell_1}
\safemath{\ellzero}{\ell_0}
\safemath{\ellinf}{\ell_\infty}
\safemath{\ellinftilde}{\ell_{\widetilde\infty}}
\safemath{\licard}{Z(\coh,\coha,\cohb)}
\safemath{\xsol}{\hat{x}}
\safemath{\xbord}{x_b}		%Solution at the border
\safemath{\xstat}{x_s}		%Solution stationary in l0 prob
\safemath{\xstatLone}{\tilde{x}_s}
\safemath{\order}{\mathcal{O}} %order notation (big O)
\safemath{\scales}{\Theta} %scales as
\safemath{\ones}{\mathbf{1}} %all ones matrix
\safemath{\zeroes}{\mathbf{0}} %all zeroes matrix
\safemath{\thlone}{\kappa(\coh,\cohb)} %treshold l1 problem
\safemath{\constoneA}{\delta} %constant in l1 theorem to save space
\safemath{\constoneB}{\epsilon} %constant in l1 theorem to save space
\safemath{\nlarge}{L}				   %num large elements
\safemath{\sumlarge}{S_\nlarge}
\safemath{\maxlarger}{P_\nlarge}	   % maximum in Gribonval and Nielsen
\safemath{\Pzero}{\textrm{P0}}	
\safemath{\Pone}{\textrm{P1}}
\safemath{\vecfir}{\vecw}			 % \vecv element of the kernel of the dictionary, \vecv=[\vecfir \vecsec]
\safemath{\vecsec}{\vecz}
\safemath{\elvecfir}{w}              % element of vecfir
\safemath{\elvecsec}{z}				 % element of vecsec
\safemath{\nlargefir}{n}
\safemath{\normout}{\gamma}
\safemath{\auxfun}{h}
\safemath{\supp}{\textrm{supp}}%support

\safemath{\indexa}{\ell}
\safemath{\indexb}{r}
\safemath{\indexc}{i}
\safemath{\indexd}{j}

\safemath{\project}{P}%projector

\usepackage{framed}

\IEEEoverridecommandlockouts
\allowdisplaybreaks % THIS ALLOWS EQUATIONS TO BREAK THE PAGE

\newcommand{\nbit}[1]{$#1$-bit}
\newcommand{\gft}{GF($2$)}
\safemath{\hsim}{\overline{h}} %Hamming similarity
\safemath{\setwz}{\{0,1\}} %set with zero
\safemath{\setwmo}{\{\pm1\}} %set with minus 1

\usepackage{algcompatible}
\usepackage{array}
\algrenewcommand\algorithmicindent{1em}
\newcolumntype{P}[1]{>{\centering\arraybackslash}p{#1}}

\begin{document}

\title{PPAC: A Versatile In-Memory Accelerator \\ for Matrix-Vector-Product-Like Operations\thanks{The 
work of OC, AGS, and CS was supported by ComSenTer, one of the JUMP centers sponsored by the semiconductor research corporation (SRC), and by SRC nCORE task 2758.004 and the US National Science Foundation (NSF) grant ECCS-1740286 under the E2CDA program. The work of MB was supported by the Cornell University Engineering Learning Initiatives (ELI).}\author{\IEEEauthorblockN{Oscar Casta\~neda, Maria Bobbett, Alexandra Gallyas-Sanhueza, and Christoph Studer} \\
\IEEEauthorblockA{\em School of Electrical and Computer Engineering, Cornell University, Ithaca, NY \\ 
e-mail: \{oc66,\,mb2567,\,ag753,\,studer\}@cornell.edu; website: http://vip.ece.cornell.edu} \\[0.12cm]
}
}

\maketitle

\begin{abstract}
Processing in memory (PIM)  moves computation into memories with the goal of improving throughput and energy-efficiency compared to traditional von Neumann-based architectures.
Most existing PIM architectures are either general-purpose but only support atomistic operations, or are specialized to accelerate a single task. 
We propose the Parallel Processor in Associative Content-addressable memory (PPAC), a novel in-memory accelerator that supports a range of matrix-vector-product (MVP)-like operations that find use in  traditional and emerging applications. 
PPAC is, for example, able to accelerate low-precision neural networks, exact/approximate hash lookups, cryptography, and forward error correction.
The fully-digital nature of PPAC enables its implementation with standard-cell-based CMOS, which facilitates automated design and portability among technology nodes. 
To demonstrate the efficacy of PPAC, we provide post-layout implementation results in 28nm CMOS for different array sizes. 
A comparison with recent digital and mixed-signal PIM accelerators reveals that PPAC is competitive in terms of throughput and energy-efficiency, while accelerating a wide range of applications and simplifying development. 
\end{abstract}

%%%

% !TEX root = 19ASAP_PPAC.tex
%%%

\section{Introduction}

Traditional von Neumann-based architectures have taken a variety of forms that trade-off flexibility with hardware efficiency.
Central processing units (CPUs) are able to compute any given task that can be expressed as a computer program. In contrast, application-specific integrated circuits (ASICs) are specialized to accelerate a single task but achieve (often significantly) higher throughputs and superior energy-efficiency. In between reside graphics processing units (GPUs) and field-programmable gate arrays (FPGAs), that are more specialized than CPUs, but typically offer higher throughput and energy-efficiency for the supported tasks. 
The ever-growing gap between computing performance and memory access times has lead today's von Neumann-based computing systems to hit a so-called ``memory wall'' \cite{memwall}, which describes the phenomenon that most of a system's bandwidth, energy, and time is consumed by memory operations.
This problem is further aggravated with the rise of applications, such as machine learning, data mining, or 5G wireless systems, where massive amounts of data need to be processed at high rates and in an energy-efficient way. 

\begin{figure}[tp]
\centering
\includegraphics[width=0.85\columnwidth]{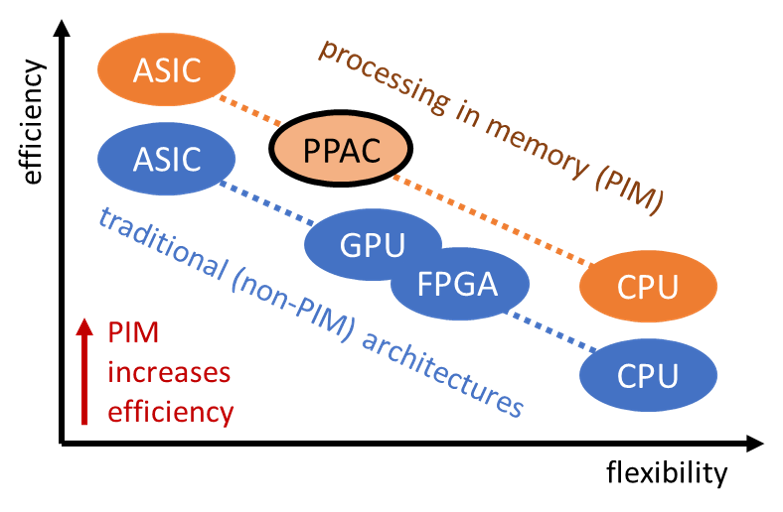}\\
\vspace{-0.25cm}
\caption{Idealized efficiency-flexibility trade-off for different hardware architectures. Processing in memory (PIM) aims at increasing throughput and energy-efficiency by moving computation into memories. The proposed Parallel Processor in Associative CAM (PPAC) is a fully-digital in-memory accelerator that supports a range of matrix-vector-product-like operations.}
\label{fig:space}
\vspace{-0.1cm}
\end{figure}

\subsection{Processing In Memory}

{Processing in memory} (PIM) is an emerging computing paradigm that promises to tear down the memory wall \cite{evo_mem_arch}. 
Put simply, PIM brings computation closer to the memories, with the objective of reducing the time and energy of memory accesses, which ultimately increases the circuit's overall efficiency (see \fref{fig:space} for an illustration).
The application of PIM to general-purpose processors has been explored recently in  \cite{compute_cache,neural_cache,acdimm}.
While such PIM-aided CPUs enable improved throughput and energy-efficiency for certain memory-intensive workloads, the supported PIM operations are typically limited to atomistic operations (such as bit-wise AND/NOR).
As a consequence, executing even slightly more complex operations (such as multi-bit additions or multiplications) requires a repeated use of the supported PIM operations; this prevents such architectures from reaching the throughput and energy-efficiency required in many of today's applications.
Hence, a number of PIM-based ASICs have been explored recently in \cite{verma,am-iscas,neurocube,drisa,brein_jssc}.
Such solutions generally excel in throughput and energy-efficiency, but have limited applicability, often accelerating a single task only. 
For example, the PIM-ASIC in \cite{verma} is designed to accelerate neural network inference using mixed-signal techniques, but suffers from effects caused by noise and process variation; this prevents its use in applications in which the least significant bit must be computed accurately (e.g., in cryptography, forward error correction, or locality-sensitive hashing).

\subsection{Contributions}
While a range of PIM-based ASICs and CPUs have been proposed in recent years,  to the best of our knowledge, no PIM-based solutions exist that simultaneously offer high flexibility and high efficiency.
To fill in this void in the trade-off space with PIM-based hardware solutions (see \fref{fig:space}), we propose a novel, versatile in-memory processor called \textit{Parallel Processor in Associative Content-addressable memory (PPAC)}, which supports a range of matrix-vector-product (MVP)-like operations.
PPAC is designed entirely in digital standard-cell-based CMOS, accelerates some of the key operations in a wide range of traditional and emerging applications, and achieves high throughput and energy-efficiency for the supported tasks.
The proposed architecture consists of a two-dimensional array of latch-based bit-cells that support two types of binary-valued operations; each row of the PPAC array is equipped with a row arithmetic-logic unit (ALU) that supports a variety of  tasks, including content-addressable memory (CAM) functionality, Hamming-distance calculation, one- and multi-bit MVPs, Galois field of two elements \gft~MVPs, and programmable logic array (PLA) functionality.
We provide post-layout implementation results in a 28\,nm CMOS technology and compare the area, throughput, and energy-efficiency to that of recent related accelerators. 

\subsection{Paper Outline}
The rest of the paper is organized as follows. 
In \fref{sec:PPACarch}, we describe the operating principle and architecture of PPAC.
In \fref{sec:apps}, we detail all operation modes and outline potential use cases.
In \fref{sec:results}, we present post-layout implementation results and compare PPAC to related accelerator designs.
We conclude in \fref{sec:conclusions}.

% !TEX root = 19ASAP_PPAC.tex
%%%

\section{PPAC: Parallel Processor in Associative CAM}
\label{sec:PPACarch}

We now describe the operating principle of PPAC and introduce its architecture.
In what follows, the terms ``word'' and ``vector'' will be used interchangeably---an $N$-bit word can also be interpreted as a binary-valued vector of dimension $N$.

\subsection{Operating Principle}

PPAC builds upon CAMs, which are memory arrays that compare all of their $M$ stored $N$-bit words $\veca_m$, $m=1,\ldots,M$, with an $N$-bit input word $\vecx$ to determine the set of stored words that match the input.
Conceptually, the functionality of a CAM can be described as a memory in which every bit-cell contains an XNOR gate to determine whether the stored value $a_{m,n}$ matches the input bit~$x_n$, $n=1,\ldots,N$. A match is then declared only if all the $N$ bits in $\veca_m$ match with the~$N$ bits of the input $\vecx$.
Mathematically, the functionality of a CAM can be expressed in terms of the \emph{Hamming distance} $h(\veca_m,\vecx)$, which indicates the number of bits in which  $\veca_m$ and~$\vecx$ differ.
A CAM declares a match between the stored word $\veca_m$ and the input word $\bmx$ if $h(\veca_m,\vecx)=0$.
As it will become useful later, one can alternatively describe a CAM's functionality using the \emph{Hamming similarity}, which we define as $\hsim(\veca_m,\vecx)=N-h(\veca_m,\vecx)$, and corresponds to the number of bits that are equal between the words $\veca_m$ and $\vecx$.
With this definition, a CAM declares a match if $\hsim(\veca_m,\vecx)=N$.
From a circuit perspective, the Hamming similarity can be computed by performing a \emph{population count} that counts the number of ones over all XNOR outputs of the CAM bit-cells of a word.

In short, PPAC builds upon a CAM that is able to compute the Hamming similarity $\hsim(\veca_m,\vecx)$ for each word $\bma_m$, $m=1,\ldots,M$, in parallel during a single clock cycle. In addition, PPAC includes (i) an additional bit-cell operator (besides the XNOR) and (ii) a simple ALU per row that enables a wide range of applications.
Since $\hsim(\veca_m,\vecx)$ is available, PPAC can implement not only a standard \emph{complete-match} CAM that declares a match whenever $\hsim(\veca_m,\vecx)=N$, but also a \emph{similarity-match} CAM that declares a match whenever the number of equal bits between $\bma_m$ and $\bmx$ meets a programmable threshold~$\delta$; i.e., $\hsim(\veca_m,\vecx) \ge \delta$.
As shown in \fref{sec:hamsim}, this similarity-match functionality finds use in different applications.

It is important to realize that with the availability of the Hamming similarity $\hsim(\veca_m,\vecx)$, PPAC can also compute an inner-product between the vectors $\veca_m$ and $\vecx$.
Assume that the entries of the $N$-dimensional binary-valued vectors $\veca_m$ and $\vecx$ are defined as follows:
If the $n$th bit has a logical high (HI) value, then the  $n$th entry represents a $+1$;
if the $n$th bit has a logical low (LO) value, then the $n$th entry represents a $-1$.
For this mapping, the inner-product between $\veca$ and $\vecx$ is
\begin{equation}
\langle\bma_m,\bmx\rangle= \sum_{n=1}^{N} a_{m,n} x_n =  2\hsim(\veca_m,\vecx)-N.
\label{eq:bindotprod}
\end{equation}
To see this, note that since $a_{m,n}, x_n \in \setwmo$, each of the partial products $a_{m,n} x_n$  is $+1$ if $a_{m,n}=x_n$ and $-1$ if $a_{m,n}\ne x_n$; this partial product can be computed with an XNOR.
If all  of the $N$ entries between $\veca_m$ and $\vecx$ differ, then $\langle\bma_m,\bmx\rangle=-N$.
Otherwise, for each bit $n$ for which $a_{m,n} = x_n$, the partial product $a_{m,n} x_n$ will change from $-1$ to $+1$, increasing the inner-product sum by $2$.
As the total number of bits that are equal between $\veca_m$ and $\vecx$ is given by $\hsim(\veca_m,\vecx)$, it follows that we can compute $\langle\bma_m,\bmx\rangle$ as in \fref{eq:bindotprod}.
Note that PPAC computes the inner-product $\langle\bma_m,\bmx\rangle$ in parallel for all the stored words~$\veca_m$, $m=1,\ldots,M$, which is exactly a $1$-bit MVP $\bA\bmx$ between the matrix~$\matA$ (whose rows are the words~$\bma_m$) and the input vector~$\vecx$. Such MVPs can be computed in a single clock cycle.

As we will show in \fref{sec:apps}, PPAC can compute multi-bit MVPs \emph{bit-serially} over several clock cycles.
Furthermore, while the XNOR gate was used to multiply $\setwmo$ entries, an AND gate can be included in each bit-cell to enable the multiplication of $\setwz$ entries.
With this AND functionality, PPAC can additionally perform (i) operations in \gft, (ii) standard unsigned and $2$'s-complement signed arithmetic, and (iii) arbitrary Boolean functions in a similar fashion to a PLA.

\subsection{Architecture Details}

\begin{figure*}[tp]
\centering
\subfigure[High-level PPAC architecture.]{\includegraphics[width=0.745\textwidth]{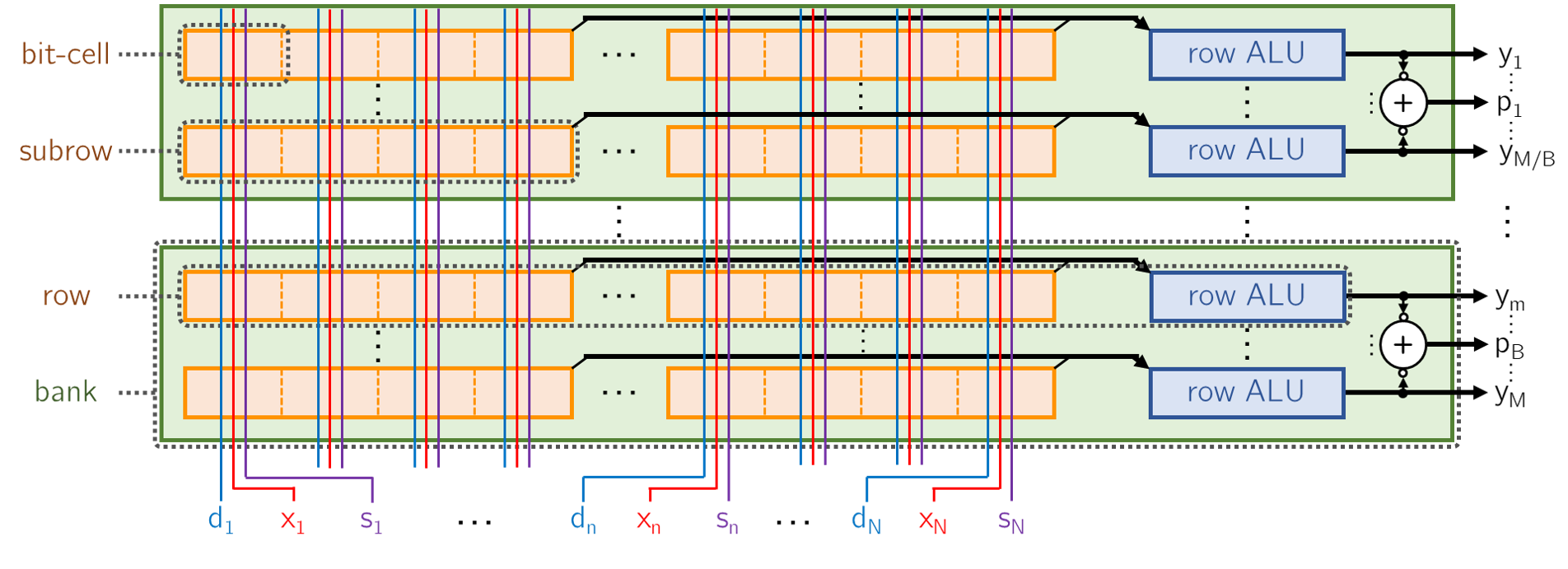}\label{fig:ppac_all}}
\vspace{-0.0cm}
\subfigure[Bit-cell and subrow details.]{\includegraphics[width=0.745\textwidth]{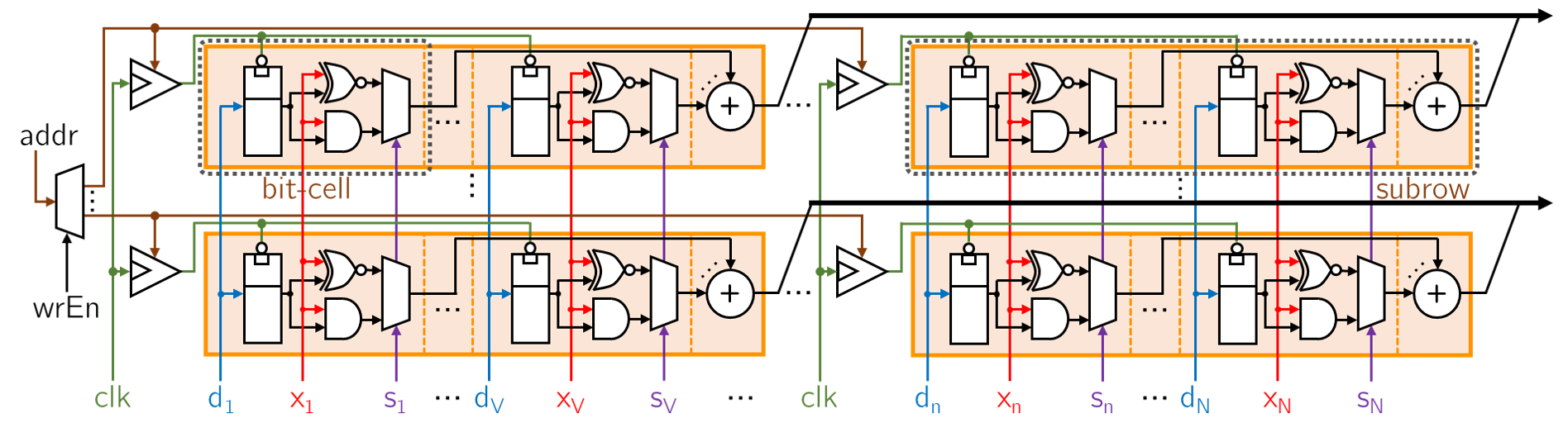}\label{fig:ppac_mem}}
\vspace{0.0cm}
\subfigure[Row ALU details. ]{\includegraphics[width=0.655\textwidth]{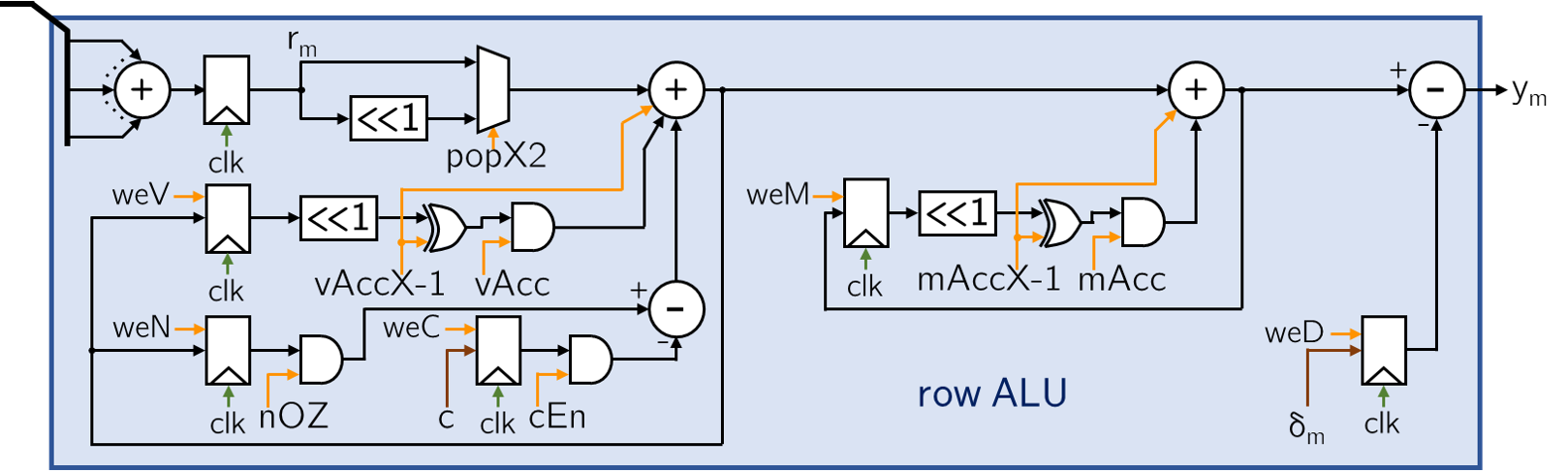}\label{fig:ppac_alu}}
\caption{Parallel Processor in Associative CAM (PPAC) architecture. (a) High-level architecture. (b) Each bit-cell includes an XNOR and an AND gate to perform bit-wise $\setwmo$ and  $\setwz$ multiplications, respectively. Writing to the bit-cell latches is accomplished using clock gates. (c) Each row of bit-cells is connected to a row ALU; the fixed-amount shifters are used to multiply the input by two; control signals are colored in orange, external data inputs in brown.}
\label{fig:PPAC}
\end{figure*}

The high-level PPAC architecture is depicted in  \fref{fig:ppac_all} and consists of multiple banks (green boxes) containing multiple rows.
Each of the $M$ PPAC rows stores an $N$-bit word in its memory (orange boxes) and is equipped with a row ALU (blue boxes).
The row ALU adds the $N$ one-bit results coming from all of the bit-cells on the row using a population count. The row population count is then used to perform different operations in the row ALU, such as Hamming-similarity or inner-product computation.
Finally, each of the $B$ banks (green boxes) contains a population count that sums up the negation of the most significant bits (MSBs) of all the row ALU's outputs. As detailed in \fref{sec:pla}, this operation enables PPAC to implement PLA functionality.

The PPAC bit-cell architecture is depicted in \fref{fig:ppac_mem}. 
All of the bit-cells corresponding to the $n$th bit position in all words $m=1,\ldots,M$ share three input signals: (i) $\mathtt{d}_n$ is the bit that will be stored in the bit-cell, (ii) $\mathtt{x}_n$ is the $n$th bit of the input word $\vecx$, and (iii) $\mathtt{s}_n$ determines if the bit-cell operator will be the XNOR or AND gate.
Each bit-cell contains a memory element (an active-low latch) that stores the input $\mathtt{d}_n$.
The bit-cells contain XNOR and AND gates to perform multiplications between the input $\mathtt{x}_n$ and the stored bit $a_{m,n}$, as well as a multiplexer, controlled by the input $\mathtt{s}_n$ that selects the bit-cell operation.
The bit-cell storage elements are written only if the address \texttt{addr} corresponding to that row and the write enable signal \texttt{wrEn} are asserted; we use clock gates to implement this functionality.
Once the memory elements are written and the control signal $\mathtt{s}_n$ has been fixed for each column, different input vectors $\vecx$ can be applied to PPAC. Then, the bit-cell operation results are passed to the row ALU, which accumulates the outputs and performs additional operations.
To improve PPAC's scalability to large arrays, each row memory is divided into $B_s$ subrows. 
Each subrow performs a population count over its $V=N/B_s$ bit-cells' results using a local adder.
With this partitioning scheme, the number of wires between each subrow and the row ALU decreases from $V$ to $\ceil{\log_2{(V+1)}}$, where~$\ceil{\cdot}$ is the ceiling function. 

The PPAC row ALU architecture is depicted in \fref{fig:ppac_alu}.
The row ALU first adds the incoming local population counts of all subrows and computes the total population count $r_m$ of the bit-cells' results for the entire row $m$.
Note that, when the XNOR operator is being used in all of the row's bit-cells, we have $r_m = \hsim(\veca_m,\vecx)$.
The result $r_m$ is then passed through two accumulators.
The first accumulator is used in applications where the vector~$\bmx$ has multi-bit entries. In this case, the MVP is carried out in a bit-serial fashion. The adder of the first accumulator also has an input to include an offset that can be used to adjust the row population count $r_m$ according to the application.
The second accumulator is used in applications where the matrix $\bA$ has multi-bit entries.
A programmable threshold $\delta_m$ is then subtracted from the output of the second accumulator to generate the row ALU's output $y_m$, whose interpretation depends on the operation mode.
In \fref{sec:apps}, we will describe how the row ALU is configured (and its output is interpreted) for each PPAC operation mode.
Note that the row ALU contains two quantities that must be stored at configuration time:
(i) The offset $c$ used to correctly interpret the row population count $r_m$ (the offset $c$ is the same for all rows for a given application) and (ii) the threshold $\delta_m$ (the threshold $\delta_m$ can be different for each row).
Finally, to increase the throughput of PPAC, we added a pipeline stage after the row population count; this increases the latency of all $1$-bit operations to two clock cycles, but a new result of a $1$-bit operation will be generated in every clock cycle.

% !TEX root = 19ASAP_PPAC.tex
%%%

\section{PPAC Operation Modes and Applications}
\label{sec:apps}

We now describe the different operating modes of the proposed PPAC and outline corresponding applications. 
In the following descriptions, we assume that all the unspecified control signals in the row ALU (cf. \fref{fig:ppac_alu}) have a value of 0; write enable (\texttt{we}) signals are set as required by the operation.

\subsection{Hamming Similarity}
\label{sec:hamsim}

In this mode, PPAC computes the Hamming similarity between the $M$ words $\bma_m$, $m=1,\ldots,M$, stored in each row  and the input word $\bmx$.
To this end, the bit-cells are configured to use the XNOR operator, so that the row population count~$r_m$ corresponds to $\hsim(\bma_m,\bmx)$.
The row ALU is configured to pass this result to PPAC's output (by setting all control signals and $\delta_m$ to $0$), so that  $y_m=\hsim(\bma_m,\bmx)$ is the Hamming similarity.

By setting $\delta_m=N$, PPAC can be used as a regular CAM. 
If all the bits of the stored word $\bma_m$ match the bits of~$\bmx$, then $r_m=N$; hence, we have $y_m=0$ and declare a match.
Otherwise, if $r_m<N$, then $y_m<0$.
Thus, a complete-match can be declared by just looking at the MSB of the output~$y_m$.
By setting $0\le \delta_m \le N$, PPAC declares a similarity-match whenever $\hsim(\bma_m,\bmx) \ge \delta_m$.
Note that PPAC performs $M$ parallel Hamming-similarity computations in each clock cycle.

In this operation mode, PPAC can be used for applications that rely on CAMs \cite{cam_apps}, including network switches and routers~\cite{cam_router}, computer caches \cite{cachecam}, and content-addressable parallel processors (CAPPs) \cite{capp,stormon}.
In this mode, PPAC can also be used for particle track reconstruction \cite{am-iscas} and for locality-sensitive hashing (LSH), which enables computationally efficient approximate nearest neighbor search \cite{andoni_lsh}.

\subsection{1-bit Matrix-Vector-Products}
\label{sec:onemvp}

In this mode, PPAC computes one MVP $\bmy=\bA\bmx$ per clock cycle, where $y_m=\langle\bma_m,\bmx\rangle$, $m=1,\ldots,M$, and $\bma_m$ and $\bmx$ are both $N$-dimensional vectors with $1$-bit entries.
We now detail how PPAC is able to support different $1$-bit number formats.  

\subsubsection{Matrix and Vector with \setwmo Entries}
In this configuration, the LO and HI logical levels are interpreted as $-1$ and $+1$, respectively, for both the matrix $\matA$ stored in PPAC and the input vector~$\vecx$.
Multiplication between a bit in $\bma_m$ (the $m$th row of $\matA$) and a bit in $\vecx$ can be computed via the bit-cell's XNOR gate. 
However, the row population count $r_m$ is an unsigned number in the range $[0,N]$.
To obtain the inner product $\langle\bma_m,\bmx\rangle$ from $r_m$, we use \fref{eq:bindotprod}, which can be implemented in the row ALU by setting $\texttt{cEn}=1$, $c=N$, and \texttt{popX2} to double the row population count (by left-shifting $r_m$ once).

\subsubsection{Matrix and Vector with \setwz Entries}
In this configuration, the LO and HI logical levels are interpreted as $0$ and $1$, respectively, for both the matrix and input vector.
Multiplication between a bit in $\veca_m$ and a bit in $\vecx$ will be $1$ only if both entries are $1$; this corresponds to using the AND gate in each bit-cell.
Hence, the row population count satisfies $r_m=\langle\bma_m,\bmx\rangle$, which can be passed directly to the row ALU output $y_m$.

\subsubsection{Matrix with \setwmo and Vector with \setwz Entries}
\label{sec:mzvo}
In this configuration, the vector $\bmx$ is expressed as $\bmx=0.5(\hat{\vecx}+\mathbf{1})$, where $\hat{\vecx}$ has $\setwmo$ entries and $\mathbf{1}$ is the all-ones vector.
Note that $\hat{\vecx}$ can be easily obtained by setting the entries of  $\vecx$ that are $0$ to $-1$; i.e., $\hat{\vecx}$ and $\vecx$ are equivalent in terms of logical  LO and HI levels.
Using \fref{eq:bindotprod}, we have the following equivalence: 
\begin{equation}
\langle\bma_m,\bmx\rangle = \hsim(\veca_m,\hat{\vecx}) + \hsim(\veca_m,\mathbf{1}) - N.
\label{eq:matpm1vec01}
\end{equation}
This requires us to compute $\hsim(\bma_m,\mathbf{1})$, which can be obtained in the Hamming-similarity mode with input vector $\mathbf{1}$.
The result of this operation is stored in the row ALU by setting \texttt{weN} to $1$.
To complete \fref{eq:matpm1vec01}, the Hamming-similarity mode is applied again, but this time with $\bmx$ (which has the same logical representation as $\hat{\bmx}$) as the input vector, and with $\texttt{nOZ}$ and $\texttt{cEn}$ set to $1$ and $c=N$.
Note that $\hsim(\veca_m,\mathbf{1})$ needs to be computed once only if the matrix~$\matA$ changes.  

\subsubsection{Matrix with \setwz and Vector with \setwmo Entries}
In this configuration,  the vector $\vecx$ is expressed as $\vecx=2\tilde{\vecx}-\mathbf{1}$, where~$\tilde{\vecx}$ has $\setwz$ entries and, as above, has the same logical LO and HI levels as $\vecx$. By noting that $\langle\bma_m,\mathbf{1}\rangle=N-\hsim(\veca_m,\mathbf{0})$, where~$\mathbf{0}$ is the all-zeros vector, we have the following equivalence:
\begin{equation}
\langle\bma_m,\vecx\rangle = 2\langle\bma_m,\tilde{\vecx}\rangle + \hsim(\veca_m,\mathbf{0}) - N.
\label{eq:mat01vecpm1}
\end{equation}
As in \fref{eq:matpm1vec01}, this requires us to  compute $\hsim(\veca_m,\mathbf{0})$, which can be obtained in the Hamming-similarity mode with input vector~$\mathbf{0}$. The result of this operation is stored in the row ALU (by setting \texttt{weN} to $1$).
One can then compute a 1-bit \setwz MVP to obtain $\langle\veca_m,\tilde{\vecx}\rangle$ for all PPAC rows $m=1,\ldots,M$, but this time with $\texttt{popX2}$, $\texttt{nOZ}$, and $\texttt{cEn}$ set to $1$, and $c=N$ to complete \fref{eq:mat01vecpm1}.
As above, $\hsim(\veca_m,\mathbf{0})$ has to be computed only if $\matA$ changes.

\nbit{1} $\setwmo$ MVPs can, for example, be used for inference of binarized neural networks \cite{bnn}.
While \nbit{1} MVPs in the other number formats might have limited applicability, they are used for multi-bit operations  as described next.

\subsection{Multi-bit Matrix-Vector-Products}
\label{sec:multimvp}

In this mode, PPAC computes MVPs $\bmy=\bA\bmx$ where the entries of $\bA$ and/or $\bmx$ have multiple bits.
All of these multi-bit operations are carried out in a bit-serial manner, which implies that MVPs are computed over multiple clock cycles.

\subsubsection{Multi-bit Vector}
Consider the case where $\matA$ has \nbit{1} entries, while the vector $\vecx$ has \nbit{L} entries. We start by writing
\begin{equation}
\bmx=\sum_{\ell=1}^{L}2^{\ell-1}\bmx_\ell,
\label{eq:multibitnum}
\end{equation}
where $\vecx_\ell$ is a \nbit{1} vector formed by the $\ell$th bit of all the entries of $\vecx$. 
This decomposition enables us to rewrite the MVP as  follows: 
\begin{equation}
\matA\vecx = \sum_{\ell=1}^{L} 2^{\ell-1}\matA\vecx_\ell.
\label{eq:multibitvec}
\end{equation}
We use PPAC's \nbit{1} MVP mode with input~$\vecx_{L}$ (the MSB of the entries of $\vecx$) to compute $\matA\vecx_{L}$. The result is stored in the first accumulator of the row ALU by setting \texttt{weV} to $1$. In the subsequent clock cycle, this value is doubled and added to $\matA\vecx_{L-1}$ by setting \texttt{vAcc} to $1$.
By repeating this operation for $\ell=L,L-1,\ldots,1$, the MVP $\bmy=\matA\vecx$ is computed \emph{bit-serially} in $L$ clock cycles.

\subsubsection{Multi-bit Matrix}
Consider the case where each entry of~$\bA$ has \nbit{K} entries.
We use the same concept as in \fref{eq:multibitvec} and we decompose $\matA=\sum_{k=1}^{K}2^{k-1}\matA_k$, where $\matA_k$ is a \nbit{1} matrix formed by the $k$th bit of all entries of $\matA$.
In contrast to the multi-bit vector case, PPAC's memory cannot be replaced to contain a different matrix $\matA_k$ every cycle.
Instead, similar to \cite{verma}, different columns of PPAC are used for different bit-significance levels, so that all $K$ bits of the entries of $\matA$ are stored in PPAC's memory.
As a result, PPAC will now contain $N/K$ different $K$-bit entries per row, instead of $N$ different $1$-bit entries per row.
To ensure that only elements from $\matA_k$ are used, the columns with different significance are configured to use the AND operator, and the corresponding entry of $\vecx$ is set to $0$, effectively nulling any contribution from these columns to the row population count $r_m$. The rest of the columns are configured according to the used number format, and $c$ in the row ALUs is set to $N/K$ for the number formats that use it, so that PPAC computes $\matA_k\vecx$ for an input $\vecx$ that has $N/K$ entries of $L$ bits.
PPAC starts by computing $\matA_{K}\vecx$ (i.e., the MVP using the most significant bit of the entries of $\matA$) and saves the result in the second accumulator of the row ALU (by setting \texttt{weM} to $1$), so that after $L$ cycles (assuming each vector entry has $L$ bits), it can double the accumulated result and add it to $\matA_{K-1}\vecx$ by setting \texttt{mAcc} to $1$.
The new accumulated result is stored in the second accumulator, which will be written again $L$ clock cycles later.
By repeating this procedure, the multi-bit MVP $\bmy=\matA\vecx$ is computed  bit-serially over~$KL$ clock cycles.

\subsubsection{Supported Number Formats}

\begin{table}[tp]
\centering
\renewcommand{\arraystretch}{1.1}
\begin{minipage}[c]{1\columnwidth}
    \centering
    \caption{$L$-bit number formats supported by PPAC}
       \label{tbl:numform}
  \begin{tabular}{@{}lccc@{}}
  \toprule
  {Name} & \texttt{uint} & \texttt{int} & \texttt{oddint} \\
   \midrule
  {LO level} & $0$ & $0$ & $-1$ \\
  {HI level} & $1$ & $1$ & $1$ \\
  {Signed?} & No & Yes & Yes \\
  \midrule
  {Min.~value} & $0$ & $-2^{L-1}$ & $-2^L+1$\\
  {Max.~value} & $2^L-1$ & $2^{L-1}-1$ & $2^L-1$\\
  \midrule
  {E.g., $L=2$} & $\{0,1,2,3\}$ & $\{-2,-1,0,1\}$ & $\{-3,-1,1,3\}$\\
  \bottomrule
  \end{tabular}
  \end{minipage}
  \end{table}

 As detailed in \fref{sec:onemvp}, PPAC is able to compute multi-bit MVPs with different number formats summarized in \fref{tbl:numform}.
For example, by mapping the logical LO level to $0$ and HI to $1$, multi-bit MVPs between unsigned numbers (\texttt{uint}) are performed.
To operate with signed numbers (\texttt{int}), we negate (in 2's complement representation) the partial products  $\bA_k\bmx_L$ (for signed multi-bit vectors) or $\bA_K\bmx$ (for signed multi-bit matrices), which are associated with the MSBs of the signed numbers in the vector $\bmx$ and matrix $\bA$, respectively.
We can configure the row ALUs to implement this behavior by setting $\texttt{vAccX-1}$ and $\texttt{mAccX-1}$ to $1$ for a signed vector or matrix, respectively.
The \texttt{oddint} number format arises from having a multi-bit number in which LO and HI get mapped to $-1$ and $+1$, respectively. Then, by applying \fref{eq:multibitnum}, \texttt{oddint} represents signed odd numbers, as illustrated in \fref{tbl:numform}. Note that \texttt{oddint} cannot represent $0$.

Low-resolution multi-bit MVPs using different number formats find widespread use in practice. 
For example, neural network inference can be executed with matrices and vectors using low-precision \texttt{int} numbers, where the threshold $\delta_m$ in the row ALU can be used as the bias term of a fully-connected (dense) layer.
A $1$-bit \texttt{oddint} matrix multiplied with a multi-bit \texttt{int} vector can be used to implement a Hadamard transform~\cite{goldstein2015stone}, which finds use in signal processing, imaging, and communication applications.

\subsection{\gft~Matrix-Vector-Products}
\label{sec:gftmvp}
In this mode, PPAC is able to perform MVPs in \gft, the finite field with two elements $\setwz$.
Multiplication in this field corresponds to an AND operation; addition corresponds to an XOR operation, which is equivalent to a simple addition  modulo-$2$. GF(2) addition can then be performed by extracting the least significant bit (LSB) of a standard integer addition.
To support MVPs in this mode, all of the columns of PPAC are set to use the AND operator in the bit-cells, and the row ALU is configured so that $y_m=r_m$. Then, the result of $\langle\bma_m,\bmx\rangle$ in \gft~can be extracted from the LSB of $y_m$. We emphasize that recent mixed-signal architectures that support MVPs, such as the ones in \cite{verma,murmann}, are unable to support this mode as the LSBs of analog additions are generally not bit-true. 

GF(2)~MVPs find widespread application in the computation of substitution boxes of encryption systems, including AES~\cite{daemen2013design}, as well as in encoding and decoding of error-correction codes, such as low-density parity-check~\cite{ldpc_bitflip} and polar codes \cite{arikan2008channel}. 

\subsection{Programmable Logic Array}
\label{sec:pla}
In this mode, each PPAC bank is able to compute a Boolean function as a sum of min-terms, similar to a PLA.
To this end, the $m$th row computes a min-term as follows:
Each PPAC column and entry of the input vector $\bmx$ correspond to a different Boolean variable $X$; note that we consider the complement $\overline{X}$  as a different Boolean variable that is associated with another column and input entry.
Then, if the Boolean variable associated with the $n$th column should appear in the min-term computed by the $m$th row, the $a_{m,n}$ bit-cell must store a logical $1$, otherwise a logical $0$.
Furthermore, all PPAC columns are set to use the AND operator, and the row ALU is configured so that $y_m=r_m-\delta_m$, where the threshold $\delta_m$ must be the number of Boolean variables that are in the $m$th row's min-term (i.e., the number of logical $1$'s stored in $\bma_m$).
By doing so, $y_m=0$ only if all of the Boolean variables in the min-term are $1$; otherwise, $y_m < 0$.
This implies that the result of the min-term of the $m$th PPAC row can be extracted from the complement of the MSB of $y_m$.
Finally, the results of all min-terms in the $b$th bank are added together using the bank adder (see~the adder in~\fref{fig:ppac_all}). If $p_b>0$, then at least one of the min-terms has a value of $1$, so the output of the Boolean function programmed in the bank is a logical $1$; otherwise, it is a logical $0$.

Note that PPAC also supports different logic structures.
For example, if we set $\delta_m=1$, then each row will be computing a max-term.
If we interpret the result of the Boolean function to be $1$ only if $p_b$ is equal to the number of programmed max-terms in the bank, PPAC effectively computes a product of max-terms.
In general, PPAC can execute a logic function with two levels: The first stage can be a multi-operand AND, OR, or majority gate (MAJ) of the Boolean inputs;  the second stage can be a multi-operand AND, OR, or MAJ of the outputs of the first stage.
With this, PPAC can be used as a  look-up table or programmed as a PLA that computes Boolean functions.
% !TEX root = 19ASAP_PPAC.tex
%%%

\section{Implementation Results}
\label{sec:results}

We now present post-layout implementation results of various PPAC array sizes in $28$\,nm CMOS and provide a comparison to existing in-memory accelerators and other related designs. 

\subsection{Post-Layout Implementation Results}

\begin{table}[tp]
\centering
\renewcommand{\arraystretch}{1.1}
\begin{minipage}[c]{1\columnwidth}
    \centering
    \caption{Post-layout implementation results for different~PPAC~array~sizes~in~28nm~CMOS}
       \label{tbl:implresultsPPAC}
  \begin{tabular}{@{}lcccc@{}}
  \toprule
  {Words $M$} & $16$ & $16$ & $256$ & $256$\\
  {Word-length $N$} & $16$ & $256$ & $16$ & $256$\\
  {Banks $B$} & $1$ & $1$ & $16$ & $16$\\
  {Subrows $B_s$} & $1$ & $16$ & $1$ & $16$\\
  \midrule
  {Area [$\mu\text{m}^2$]} & 14\,161 & 72\,590 & 185\,283 & 783\,240 \\
  {Density [\%]} & 75.77 & 70.45 & 72.52 & 72.13 \\
  {Cell area [kGE]} & 17 & 81 & 213 & 897 \\
  {Max.~clock freq.~[GHz]} & 1.116 & 0.979 & 0.824 & 0.703 \\
  {Power [mW]} & 6.64 & 45.60 & 78.65 & 381.43 \\
  \midrule
  {Peak throughput [TOP/s]} & 0.55 & 8.01 & 6.54 & 91.99 \\
  {Energy-eff.~[fJ/OP]} & 12.00 & 5.69 & 12.03 & 4.15 \\  
  \bottomrule
  \end{tabular}
  \end{minipage}
  \end{table}
  
\begin{figure}[tp]
\centering
\includegraphics[width=0.9\columnwidth]{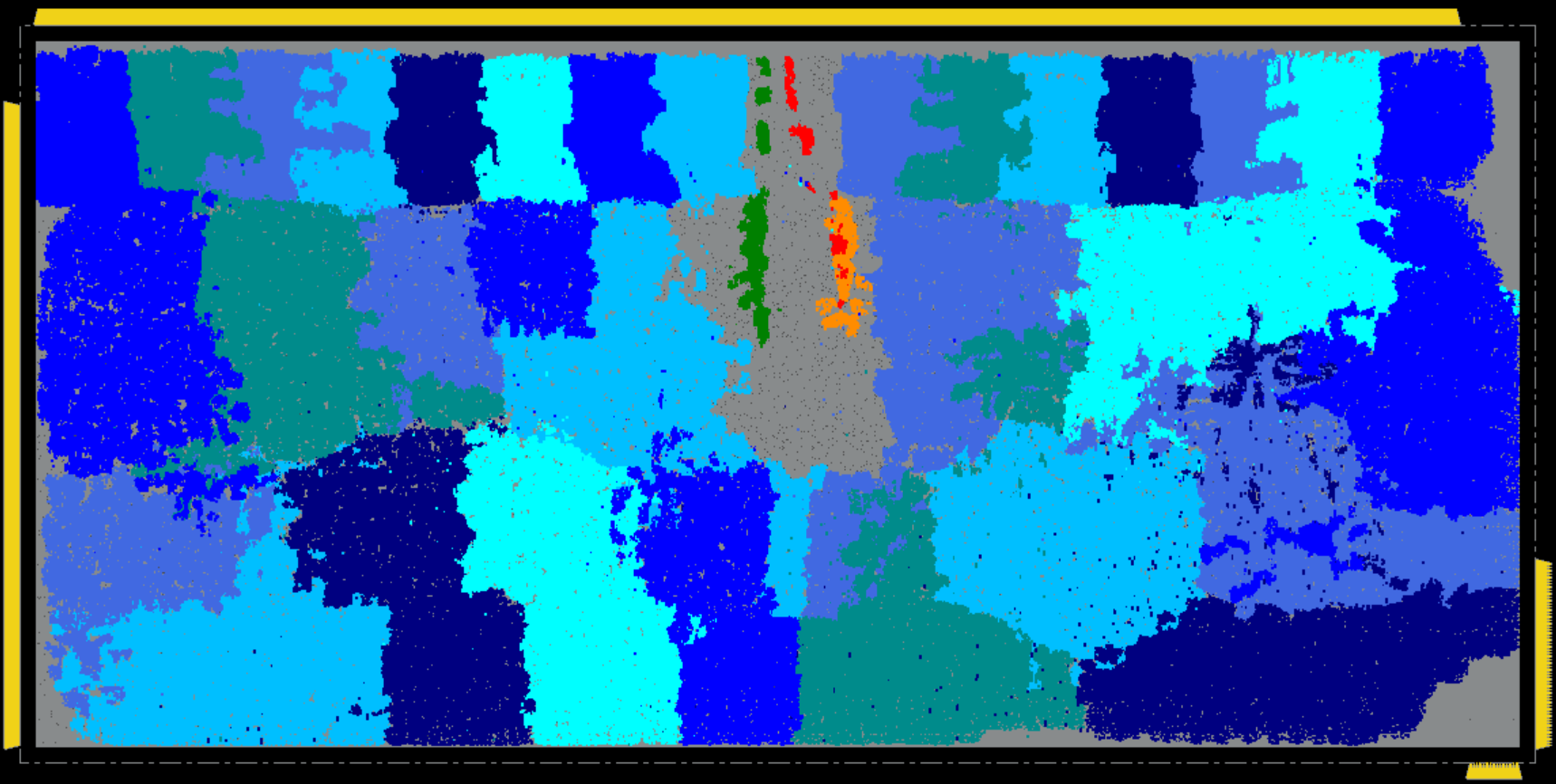}
\caption{Layout of the $256\times256$ PPAC with $B=B_s=16$. All banks but one are colored using different shades of blue. For the gray bank, one row is shown in green, while the row memory and row ALU of another row are shown in orange and red, respectively.}
\label{fig:layout}
\end{figure}

We have implemented four different $M\times N$ PPAC arrays in $28$\,nm CMOS.
All of these PPAC implementations have banks formed by $16$ rows, each with $V=16$ bit-cells per subrow, and a row ALU that supports multi-bit operations with~$L$ and~$K$ up to $4$ bits. 
In \fref{tbl:implresultsPPAC}, we summarize our post-layout implementation results;  the CAD-generated layout of the $256\times256$ PPAC design is shown in \fref{fig:layout}.
The throughput is measured in operations (OP) per second, where we count both \nbit{1} multiplications and \nbit{1} additions as one OP each. Since each PPAC row performs an inner product between two $N$-dimensional $1$-bit vectors, an $M\times N$ PPAC performs $M(2N\!-\!1)$ OP per clock cycle.
Even if the clock frequency decreases as PPAC's dimensions increase, the overall throughput increases up to $92$\,TOP/s for the $256\times256$ array; this occurs due to the massive parallelism of our design.
We also observe that increasing the number of words $M$ results in a higher area and power consumption than increasing the number of bits per word $N$ by the same factor. This behavior is due to the fact that adding a new row implies including a new row ALU, whose area can be comparable to that of the row memory (cf.~\fref{fig:layout}). In contrast, increasing the number of bits per word $N$ mainly modifies the datapath width of an existing row ALU, which scales only logarithmically in $N$, improving the energy-efficiency of the $256\times256$ PPAC to 4.15\,fJ/OP.

\begin{table}[tp]
\centering
\renewcommand{\arraystretch}{1.1}
\begin{minipage}[c]{1\columnwidth}
  \centering
  
  \caption{Throughput, power, and energy-efficiency for different applications with a $256\times256$ PPAC array in 28nm CMOS}
  \label{tbl:powerPPAC}
  \begin{tabular}{@{}lccc@{}}
  \toprule
  \multirow{2}{*}{Operation mode} & {Throughput} & {Power} & {Energy-eff.}\\
  { } & {[GMVP/s]} & {[mW]} & {[pJ/MVP]}\\
   \midrule
  {Hamming similarity} & 0.703 & 478 & 680 \\
  {\nbit{1} $\setwmo$ MVP} & 0.703 & 498 & 709 \\
  {\nbit{4} $\setwz$ MVP} & 0.044 & 226 & 5\,137 \\
  {\gft~MVP} & 0.703 & 353 & 502 \\
  {PLA} & 0.703 & 352 & 501 \\
  \bottomrule
  \end{tabular}
  \end{minipage}
  \end{table}
  
In \fref{tbl:powerPPAC}, we summarize the throughput, power, and energy-efficiency for the different operation modes executed on a $256\times256$ PPAC. 
Throughput and energy-efficiency are measured in terms of MVPs, where for the Hamming-similarity mode, an MVP corresponds to the computation of $M=256$ Hamming similarities; for the PLA mode, an MVP computes~$B=16$ distinct Boolean functions.
To extract power estimates, we used Cadence Innovus and stimuli-based post-layout simulations at 0.9\,V and 25\,C$^\circ$ in the typical-typical process corner.
In our simulations, we first load a randomly-generated matrix~$\matA$ into PPAC's memory, and then apply $100$ random input vectors~$\bmx$ for the \nbit{1} operations, while for the \nbit{4} \setwz~MVP case, we execute $100$ different MVPs.
We simulate the dynamic and static power consumption of PPAC only while performing computations (i.e., we exclude the power consumption of initializing the matrix~$\matA$), as this is the envisioned use case for PPAC---applications in which the matrix~$\matA$ remains largely static but the input vectors $\vecx$ change at a fast rate.
From \fref{tbl:powerPPAC}, we observe that operations that use the XNOR operator (i.e., Hamming similarity and \nbit{1} $\setwmo$ MVP) exhibit higher power consumption than tasks relying on the AND operation; this is because the switching activity at the output of XNOR gates is, in general, higher than that of AND gates.

\begin{table*}[tp]
\centering
\renewcommand{\arraystretch}{1.1}
\begin{minipage}[c]{2\columnwidth}
    \begin{center}
    \caption{Comparison with existing binarized neural network (BNN) accelerator designs}
       \label{tbl:comp}
  \begin{tabular}{@{}l|ccc|ccc|cc|cc@{}}
  \toprule
  \multirow{2}{*}{Design} & \multirow{2}{*}{PIM?} & {Mixed} & \multirow{2}{*}{Implementation} & {Technology} & {Supply} & {Area} & {Peak TP} & {Energy-eff.} & {Peak TP$^\textit{a}$} & {Energy-eff.$^\textit{a}$}\\
  & & {signal?} &  & {[nm]} & {[V]} & {[$\text{mm}^2$]} & {[GOP/s]} & {[TOP/s/W]} & {[GOP/s]} & {[TOP/s/W]}\\
  \midrule
  {PPAC} & {yes} & {no} & {layout} & {28} & {0.9} & {0.78} & {91\,994} & {184} & {91\,994} & {184} \\
 {CIMA \cite{verma}} & {yes} & {yes} & {silicon} & {65} & {1.2} & {8.56} & {4\,720} & {152} & {10\,957} & {1\,456} \\
  {Bankman \emph{et al.} \cite{murmann}} & {no} & {yes} & {silicon} & {28} & {0.8} & {5.95} & {--} & {532} & {--} & {420} \\
  {BRein \cite{brein_jssc}} & {yes} & {no} & {silicon} & {65} & {1.0} & {3.9} & {1.38} & {2.3} & {3.2} & {15} \\
  {UNPU \cite{unpu_isscc}} & {no} & {no} & {silicon} & {65} & {1.1} & {16} & {7\,372} & {46.7$^\textit{b}$} & {17\,114} & {376} \\
  {XNE \cite{xne}} & {no} & {no} & {layout} & {22} & {0.8} & {0.016} & {108} & {112} & {84.7} & {54.6} \\
  \bottomrule
  \end{tabular}
  \end{center}
\hspace{0.05cm} {\footnotesize $^\textit{a}$Technology scaling to 28\,nm CMOS at $V_\text{dd}=0.9\,\text{V}$ assuming standard scaling rules $A\sim1/\ell^2$, $t_\text{pd}\sim1/\ell$, and $P_\text{dyn}\sim1/(V_\ell^2\ell)$.\\\phantom{ll}$^\textit{b}$Number reported in \cite[Fig. 13]{unpu_isscc}; note that the peak TP (7\,372 GOP/s) divided by the reported power consumption (297 mW) yields 24.8 TOP/s/W.}

  \end{minipage}
  \end{table*}

\subsection{Comparison with Existing Accelerators}

In \fref{tbl:comp}, we compare the $256\times256$ PPAC with existing hardware accelerators that have been specialized for binarized neural network (BNN) inference and support fully-connected layers~\cite{verma,murmann,brein_jssc,unpu_isscc,xne}.
We compare against these designs as their operation closely resembles that of PPAC's $1$-bit \setwmo MVP operation mode. In fact, all of the considered designs count $1$-bit products and additions as one operation (OP) each---an inner product between two $N$-dimensional $1$-bit vectors is $2N$ OPs.
The designs in \cite{verma,brein_jssc} are PIM accelerators in which part of the computation is carried out within the bit-cells;  the designs in~\cite{verma,murmann} rely on mixed-signal techniques to compute MVPs.

By considering technology scaling, we see that the energy~efficiency (in terms of TOP/s/W) of PPAC is comparable to that of the two fully-digital designs in \cite{unpu_isscc,xne} but $7.9\times$ and $2.3\times$ lower than that of the mixed-signal designs in \cite{verma} and \cite{murmann}, respectively, where the latter is implemented in a comparable technology node as PPAC.
As noted in \fref{sec:gftmvp}, mixed-signal designs are particularly useful for tasks that are resilient to noise or process variation, such as neural network inference. However, mixed-signal designs cause issues in applications that require bit-true results, such as addition in \gft, which requires the LSB of an integer addition to be exact.

We also see that PPAC achieves the highest peak throughput among the considered designs, which is due to its massive parallelism.
We emphasize, however, that PPAC's performance was extracted from post-layout simulations, whereas all the other designs, except that in~\cite{xne}, are silicon-proven.
Furthermore, all other designs not only execute $1$-bit MVPs, but they also include other operations that are required to implement BNN inference, such as activation functions and batch normalization.
PPAC, in contrast, is unable to completely execute BNN inference, but is able to execute a $256\times256$ MVP followed by adding a bias vector, which is a large portion of the operations required to process a fully-connected BNN layer. 
As a result, the reported throughput and energy-efficiency for PPAC are optimistic.

We would like to reiterate that PPAC is a massively-parallel PIM engine that can be used for a number of different MVP-like operations, where $1$-bit MVP is just one of them.
As such, the main purpose of the comparison in \fref{tbl:comp} is to demonstrate that PPAC's $1$-bit $\setwmo$ MVP operation mode holds promise with an energy-efficiency that is comparable to that of other accelerators.
While the hardware designs in~\cite{murmann,brein_jssc,xne} are specialized to carry out $1$-bit MVPs and the designs in~\cite{verma,unpu_isscc} to execute multi-bit MVPs for neural network inference, PPAC is programmable to perform not only these operations, but also \gft~MVPs, Hamming-similarity computations, and PLA or CAM functionality, opening up its use  in a wide range of applications.
In this sense, PPAC is similar to the work in \cite{compute_cache}, where PIM is used to accelerate multiple applications, such as database query processing, cryptographic kernels, and in-memory checkpointing.
A fair comparison to~\cite{compute_cache} is, however, difficult as it considers a complete system---PPAC would need to be integrated into a system for a fair comparison.
We note, however, that if the method in \cite{compute_cache} is used to compute MVPs, an element-wise multiplication between two vectors whose entries are $L$-bit requires $L^2+5L-2$ clock cycles~\cite{neural_cache}, which is a total of $34$ clock cycles for $4$-bit numbers. Then, the reduction (via sum) of an $N$-dimensional vector with $L$-bits per entry requires $\mathcal{O}\!\left(L\log_2(N)\right)$ clock cycles, which is at least $64$ clock cycles for a $256$-dimensional vector with $8$-bit entries (as the product of two $4$-bit numbers results in $8$-bit). Hence, an inner product between two  $4$-bit vectors with $256$ entries requires at least $98$ clock cycles---PPAC requires only $16$ clock cycles for the same operation.
This significant difference in the number of clock cycles is caused by the fact that the design in~\cite{neural_cache} is geared towards data-centric applications in which element-wise operations are performed between high-dimensional vectors to increase parallelism. PPAC aims at accelerating a wide range of MVP-like operations, which is why we included dedicated hardware (such as the row pop-count) to speed up element-wise vector multiplication and vector sum-reduction.
  
% !TEX root = 19ASAP_PPAC.tex
%%%

\section{Conclusions}
\label{sec:conclusions}

We have developed a novel, all-digital in-memory accelerator we call \emph{Parallel Processor in Associative CAM} (PPAC).  
PPAC accelerates a variety of matrix-vector-product-like operations with different number formats in a massively-parallel manner. 
We have provided post-layout implementation results in a 28nm CMOS technology for four different array sizes, which demonstrate that a $256\times256$ PPAC array achieves $92$\,TOP/s at an energy efficiency of $4.15$\,fJ/OP.
Our  comparison with recent digital and mixed-signal PIM and non-PIM accelerators has revealed that  PPAC can be competitive in terms of throughput and energy-efficiency while maintaining high flexibility.

We emphasize that the all-digital nature of PPAC has numerous practical advantages over existing mixed-signal PIM designs.
First, PPAC can be implemented using automated CAD tools with conventional standard-cell libraries and fabricated in standard CMOS technologies.
Second, PPAC is written in RTL with Verilog, is highly parametrizable (in terms of array size, banking, supported operation modes, etc.), and can easily be migrated to other technology nodes.
Third, PPAC's all-digital nature renders it robust to process variations and noise, facilitates in-silicon testing, and its clock frequency and supply voltage can be aggressively scaled to either increase throughput or improve energy-efficiency.

There are numerous avenues for future work.
The design of semi-custom bit-cells (e.g., by fusing latches with logic) has the potential to significantly reduce area and power consumption, possibly closing the efficiency gap to mixed-signal PIM accelerators.
Furthermore, guided cell placement and routing may yield higher bit-cell density and hence, potentially reduce area as well as mitigate interconnect congestions and energy. Finally, integrating PPAC into a processor either as an accelerator or compute cache is an interesting open research direction.

%%%

\balance
\bibliographystyle{IEEEtran}
\bibliography{bib}
\balance

\end{document}